\documentclass[aps,prl,superscriptaddress,twocolumn,showpacs,amsmath,amssymb,floatfix]{revtex4}

\usepackage{graphicx}
\usepackage{dcolumn}
\usepackage{bm}

\newcommand{\thetamin}{\theta_{min} \,}

\newcommand{\kthreepid}{K_{L}\to \pi^{0}\pi^{0}_{D}\pi^{0}_{D} \,}
\newcommand{\ktwopid}{K_{L}\to \pi^{0}_{D}\pi^{0}_{D} \,}
\newcommand{\meegg}{m_{ee\gamma\gamma} \,}
\newcommand{\mgg}{m_{\gamma\gamma} \,}

\newcommand{\kleegg}{K_{L}\to e^+e^-\gamma\gamma \,}
\newcommand{\klpiee}{K_{L}\to \pi^{0}e^+e^- \,}
\newcommand{\kltwopid}{K_{L}\to \pi^0\pi^{0}_{D} \,}

\newcommand{\kpmzero}{K_{L}\to \pi^+\pi^-\pi^{0} \,}
\newcommand{\kpenu}{K_{L}\to \pi^{\pm}e^{\mp} \nu \,}

\newcommand{\pitoeeg}{$\pi^0 \to e^+e^-\gamma$}

\newcommand{\UAz}{University of Arizona, Tucson, Arizona 85721}
\newcommand{\UCLA}{University of California at Los Angeles, Los Angeles,
                    California 90095}
\newcommand{\UCSD}{University of California at San Diego, La Jolla,
                   California 92093}
\newcommand{\Campinas}{Universidade Estadual de Campinas, Campinas,
                       Brazil 13083-970}
\newcommand{\EFI}{The Enrico Fermi Institute, The University of Chicago,
                  Chicago, Illinois 60637}
\newcommand{\UB}{University of Colorado, Boulder, Colorado 80309}
\newcommand{\ELM}{Elmhurst College, Elmhurst, Illinois 60126}
\newcommand{\FNAL}{Fermi National Accelerator Laboratory,
                   Batavia, Illinois 60510}
\newcommand{\Osaka}{Osaka University, Toyonaka, Osaka 560-0043 Japan}
\newcommand{\Rice}{Rice University, Houston, Texas 77005}
\newcommand{\SaoPaolo}{Universidade de Sao Paolo, Sao Paolo, Brazil 05315-970}
\newcommand{\UVa}{The Department of Physics and Institute of Nuclear and
                  Particle Physics, University of Virginia,
                  Charlottesville, Virginia 22901}
\newcommand{\UW}{University of Wisconsin, Madison, Wisconsin 53706}

\begin{document}


\title{Search for the Rare Decay $\klpiee$}

\affiliation{\UAz}
\affiliation{\UCLA}
\affiliation{\UCSD}
\affiliation{\Campinas}
\affiliation{\EFI}
\affiliation{\UB}
\affiliation{\ELM}
\affiliation{\FNAL}
\affiliation{\Osaka}
\affiliation{\Rice}
\affiliation{\SaoPaolo}
\affiliation{\UVa}
\affiliation{\UW}

\author{A.~Alavi-Harati}  \affiliation{\UW}
\author{T.~Alexopoulos}   \affiliation{\UW}
\author{M.~Arenton}       \affiliation{\UVa}
\author{R.F.~Barbosa}     \affiliation{\SaoPaolo}
\author{A.R.~Barker}      \affiliation{\UB}
\author{M.~Barrio}        \affiliation{\EFI}
\author{L.~Bellantoni}    \affiliation{\FNAL}
\author{A.~Bellavance}    \affiliation{\Rice}
\author{E.~Blucher}       \affiliation{\EFI}
\author{G.J.~Bock}        \affiliation{\FNAL}
\author{C.~Bown}          \affiliation{\EFI}
\author{S.~Bright}        \affiliation{\EFI}
\author{E.~Cheu}          \affiliation{\UAz}
\author{R.~Coleman}       \affiliation{\FNAL}
\author{M.D.~Corcoran$^{\dagger}$}     \affiliation{\Rice}
\author{B.~Cox}           \affiliation{\UVa}
\author{A.R.~Erwin}       \affiliation{\UW}
\author{C.O.~Escobar}     \affiliation{\Campinas}
\author{R.~Ford}          \affiliation{\FNAL}
\author{A.~Glazov}        \affiliation{\EFI}
\author{A.~Golossanov}    \affiliation{\UVa}
\author{P. Gouffon}       \affiliation{\SaoPaolo}
\author{J.~Graham}        \affiliation{\EFI}
\author{J.~Hamm}          \affiliation{\UAz}
\author{K.~Hanagaki}      \affiliation{\Osaka}
\author{Y.B.~Hsiung}      \affiliation{\FNAL}
\author{H.~Huang}         \affiliation{\UB}
\author{V.~Jejer}         \affiliation{\UVa}
\author{D.A.~Jensen}      \affiliation{\FNAL}
\author{R.~Kessler}       \affiliation{\EFI}
\author{H.G.E.~Kobrak}    \affiliation{\UCSD}
\author{K.~Kotera}        \affiliation{\Osaka}
\author{J.~LaDue}         \affiliation{\UB}
\author{N.~Lai}           \affiliation{\EFI}
\author{A.~Ledovskoy}     \affiliation{\UVa}
\author{P.L.~McBride}     \affiliation{\FNAL}

\author{E.~Monnier}
   \altaffiliation[Permanent address ]{C.P.P. Marseille/C.N.R.S., France}
   \affiliation{\EFI}

\author{K.S.~Nelson}     \affiliation{\UVa}
\author{H.~Nguyen}       \affiliation{\FNAL}
\author{H.~Ping}         \affiliation{\UW}
\author{V.~Prasad}       \affiliation{\EFI}
\author{X.R.~Qi}         \affiliation{\FNAL}
\author{B.~Quinn}        \affiliation{\EFI}
\author{E.J.~Ramberg}    \affiliation{\FNAL}
\author{R.E.~Ray}        \affiliation{\FNAL}
\author{M.~Ronquest}     \affiliation{\UVa}
\author{E. Santos}       \affiliation{\SaoPaolo}
\author{K.~Senyo}        \affiliation{\Osaka}
\author{P.~Shanahan}     \affiliation{\FNAL}
\author{J.~Shields}      \affiliation{\UVa}
\author{W.~Slater}       \affiliation{\UCLA}
\author{D.E.~Smith}      \affiliation{\UVa}
\author{N.~Solomey}      \affiliation{\EFI}
\author{E.C.~Swallow}    \affiliation{\EFI}\affiliation{\ELM}
\author{S.A.~Taegar}     \affiliation{\UAz}
\author{R.J.~Tesarek}    \affiliation{\FNAL}
\author{P.A.~Toale}      \affiliation{\UB}
\author{R.~Tschirhart}   \affiliation{\FNAL}
\author{C. Velissaris}   \affiliation{\UW}
\author{Y.W.~Wah}        \affiliation{\EFI}
\author{J.~Wang}         \affiliation{\UAz}
\author{H.B.~White}      \affiliation{\FNAL}
\author{J.~Whitmore}     \affiliation{\FNAL}
\author{M.~Wilking}      \affiliation{\UB}
\author{B.~Winstein}     \affiliation{\EFI}
\author{R.~Winston}      \affiliation{\EFI}
\author{E.T.~Worcester}  \affiliation{\EFI}
\author{T.~Yamanaka}     \affiliation{\Osaka}
\author{R.F.~Zukanovich} \affiliation{\SaoPaolo}

\begin{abstract}

The KTeV/E799 experiment at Fermilab has searched for the 
rare kaon decay $\klpiee$.  This mode is 
expected to have a significant CP violating component.  The measurement of its 
branching ratio could support the Standard Model or could indicate the 
existence of new physics. This Letter reports new results from the 
1999-2000 data set.
One event is observed with an expected background at
0.99 $\pm$ 0.35 events.  We set a limit on the branching ratio of 
3.5 $\times 10^{-10}$ at the 90\% confidence level.
Combining with the previous result based on the dataset taken 
in 1997 yields the final KTeV result: 
BR($\klpiee$) $<$ 2.8 $\times 10^{-10}$ at 90\% C.L.

\end{abstract}

\maketitle

The decay $K_L \to \pi^0 e^+e^-$ has long been studied in the context
of Standard Model CP violation (CPV) and has more recently been of interest
in certain new physics scenarios. 

In the Standard Model, there are direct and indirect CPV contributions to
the amplitude, plus an interference term~\cite{leo1,old2,deip98}.
The indirect component is known from the measurement~\cite{na48talk} of
BR($K_S \to \pi^0 e^+e^-$) and appears to dominate.
The direct component has been estimated to be about 3 to 6 $\times 10^{-12}$,
and the two CPV contributions together give
BR($K_L \to \pi^0 e^+e^-)_{CPV}$ in the range 
8 to 45 $\times 10^{-12}$.  There is also a CP conserving amplitude
through  $\pi^0\gamma^* \gamma^*$ states which can be determined
from measurements of $K_L \to \pi^0 \gamma \gamma$ \cite{elliott},\cite{leo2}.
In recent work, Buchalla, D'Ambrosio, and Isidori \cite{newbuchalla}
argue that the CP-conserving contribution is negligible. They 
predict a Standard Model branching ratio 
$BR(K_L \rightarrow \pi^0 e^+e^-) \sim 3\times 10^{-11}$,
dominated by CPV, with a 40\% contribution from direct CPV, through
the interference term.

Observation of $K_L \to \pi^0 e^+e^-$ at rates substantially higher than
 Standard Model expectations 
would signal new physics. In a large class of SUSY models,
a branching ratio enhancement of up to five times the Standard Model
expectation is considered likely~\cite{old4}, but values as high as
10$^{-10}$ are not entirely ruled out.  The existing experimental
limit~\cite{us} has been used to constrain squark masses~\cite{Guido}
and SUSY contributions~\cite{Messina} to the charge asymmetry in 
$K^\pm \to \pi^\pm \ell^+\ell^-$.  The implications of a specific
model with extra dimensions for $K_L \to \pi^0 e^+e^-$ 
and related processes have been investigated in \cite{extradimensions}.   

The existing experimental upper limit on BR($K_L \to \pi^0 e^+e^-$) of
5.1 $\times 10^{-10}$ at the 90\% confidence level (CL) is based on the 1997
KTeV dataset.  In this Letter we present an improved limit based on data 
collected during 1999-2000.

At KTeV, 800 GeV/c protons from the Tevatron were directed onto a 
BeO target 
to create two parallel $K_L$ beams. The beams entered a 65m long vacuum tank,
which defines the fiducial volume for accepted decays. 
Charged particles were detected by two pairs of drift chambers
separated by an analysis magnet providing a transverse momentum 
kick of 0.150 GeV/c.   Photon vetoes positioned around the 
vacuum decay region and
the spectrometer vetoed particles escaping the drift chambers.
The KTeV detector is further described in~\cite{detector}.

Powerful discrimination against charged pions, which could fake electrons, was 
provided by a set of transition radiation detectors (TRDs) behind the 
drift chambers.  Each
of the eight planes was composed of a polypropylene felt radiator paired with a
double-plane multiwire proportional chamber containing an 80\%-20\% admixture 
of Xenon and CO$_2$.  TRD cuts resulted in a pion rejection factor of 
about 50:1, as measured in a sample of $\kpenu$ decays.  
These cuts were over 94\% efficient for electrons.  A more detailed
description of the TRD may be found in~\cite{greg_thesis}. 

Downstream of the TRDs were the trigger hodoscopes.  
The trigger required hits in the hodoscope planes and the spectrometer 
consistent with the passage of two oppositely charged particles.
The trigger hodoscopes were followed by the CsI electromagnetic 
calorimeter~\cite{calorimeter}, which had an energy resolution 
$\sigma(E)/E = 0.45\% \oplus 2\% /\sqrt{E(GeV)}$.
Electrons were identified by requiring the 
ratio of the energy measured in the calorimeter (E) to the momentum as
measured in the spectrometer (p) to be consistent with one; this cut
rejected about 99.5\% of charged pions.

A detailed package of Monte Carlo simulation routines was used to study
detector geometry and performance, as 
well as various trigger and analysis selection criteria.
The programs were also used to simulate background events and tailor cuts
to optimize the signal to background ratio.  

The $\klpiee$ final state consists of two photons, which come from the 
$\pi^0$ decay, and two electrons.  
$\klpiee$ candidates exhibit the following signature:
two tracks of opposite charge originating from a common vertex, and 
depositing all of their 
energy in the calorimeter; and two other 
clusters in the calorimeter, which, when taken as photons originating from
the vertex, have a mass consistent with the $\pi^0$ mass. 

$\kltwopid$ events, 
where $\pi^0_D$ indicates the pion Dalitz decay \pitoeeg, are
used to measure the $K_L$ flux and normalize the acceptance calculation.
This mode has a signature similar to $\klpiee$, with the addition of a photon. 

Recorded $\klpiee$ and $\kltwopid$  events satisfied the following
trigger requirements.  There must have been at least two separate
track candidates in each drift chamber plane.  There must not have
been hadronic showers in the calorimeter, and the event must have
deposited little energy in the photon vetoes.  There must have been a
minimum number of clusters in the calorimeter with energy greater than
1 GeV,  as determined by the hardware cluster counting
system~\cite{colin_bown}.  For  $\klpiee$, this number was four
clusters and for $\kltwopid$ it was five.

In the offline event reconstruction and analysis, events are 
required to satisfy further selection criteria.  
The charged tracks must point to calorimeter clusters.
To identify these tracks as electrons, the ratio of the energy of the
matched cluster as measured in the CsI (E) to track momentum as measured by
the drift chambers (p) must lie in the range 0.95 $< E/p <$ 1.05.  The track
positions must have sufficient clearance from the CsI edges. The decay 
vertex ($Z_{vtx}$)
has to be within the vacuum decay volume: 96 m $< Z_{vtx} <$ 158 m.  
The recontructed kaon 
momentum is required to be  between 20.3 and 216 GeV/c.  
Tracks are required to be well separated (greater than 1 cm apart at the first
drift chamber) and the opening angle between the two tracks 
has to be larger than 2.25 mrad
in the lab frame.

Further selection cuts for the $\kltwopid$ sample included 
requirements on the invariant masses of the $e^+e^-\gamma$, $\gamma\gamma$,
and $e^+e^- \gamma \gamma \gamma$ combinations, and on the momentum transverse
to the $K_L$ flight direction,
$p_\perp$.  A well-reconstructed kaon should have a $p_\perp$ close to zero. 
Using the calculated acceptance and 
known branching ratio for $\kltwopid$ decays, the total number 
of $K_L$ decays in the data sample is 
(349.0 $\pm$ 2.8$_{\rm stat}$ $\pm$ 21.6$_{\rm syst}$ $\pm$ 11.8$_{\rm BR}$) 
$\times$ 10$^9$.

Several backgrounds with the $e^+e^-\gamma\gamma$ final state
exist and can mimic the $\klpiee$ signal.  
The first source of background is $\kpmzero$ where both charged
pions shower in the calorimeter and appear to be electrons.
To remove this background, the mass of the event, under the 
hypothesis that the tracks were pions, is required to exceed 520 MeV/c$^2$.

The second source of background is $K_L \to  \pi^0 \pi^0$ and
$K_L \to \pi^0 \pi^0 \pi^0$ with one or two Dalitz decays of a $\pi^0$
and with one or more photons undetected. 
To ensure that all $K_L$ decay products are observed,
$p_\perp^2$ $<$ 1000 (MeV/c)$^2$ is required.
Additional background events of this type are removed by requiring that
the invariant mass of the two electrons, $m_{ee}$, exceed 140 MeV/c$^2$. 
However, there are some backgrounds involving two $\pi^0_D$ decays in which 
only one electron and one positron are reconstructed with a high mass.  
These events might also include coincident accidental activity.
These background events are rejected by requiring that $m_{ee}$ be less than
362.7 MeV/c$^2$. 

The third source of background is $\kpenu$ 
where the pion fakes an electron by showering in the calorimeter,
and photons are radiated by the electron or are accidentals.
This background is rejected by examining the 
response of the TRDs for both tracks.

After these cuts are applied, the single largest remaining
background is the radiative Dalitz decay $\kleegg$
with invariant mass of the two photons, $m_{\gamma\gamma}$, 
consistent with the $\pi^0$ mass.  
These events come from both internal and external 
bremsstrahlung; both  contributions were studied in~\cite{kleegg}.

In Figure 1, $\mgg$ is plotted against the invariant 
mass of the four-particle system, $\meegg$.
The $\mgg$ is determined under the assumption that the photons came from the
charged vertex, while $\meegg$ is calculated using the ``neutral vertex",
found by applying the $\pi^0$ mass constraint to the  
photon energies and positions in the calorimeter. 
Better $\meegg$ resolution is achieved for the signal
Monte Carlo using the neutral vertex, but this procedure gives the 
incorrect mass 
for the $\kleegg$ background, resulting in the diagonal swath in Figure 1.

\begin{figure}
\scalebox{0.9}{\includegraphics{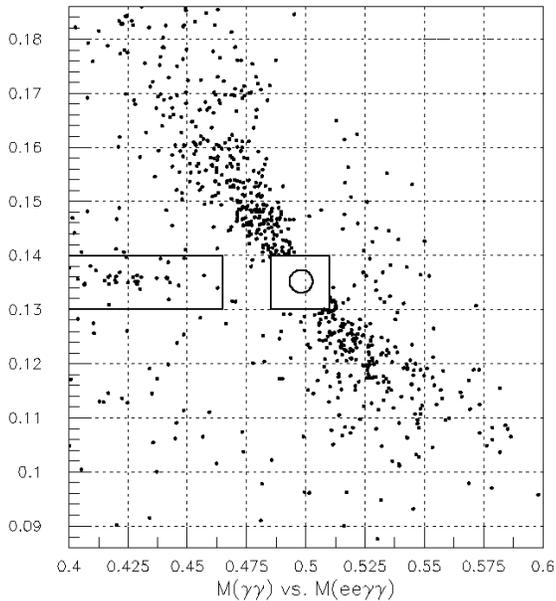}}
\caption{\label{fig:eegg_data} 
$\mgg$ (charged vertex) vs. $\meegg$ (neutral vertex) for the data
after all cuts have been applied except for the phase space cuts.
The regions appearing in the figure
are discussed in the text, and signal events in the center box have 
not been plotted.  Masses are in GeV/c$^2$.}
\end{figure} 

There are several distinctive regions in the $\mgg$ vs. $\meegg$ plane.  
In order to minimize human bias in the determination of the selection criteria,
a blind analysis was performed.  The 
box was the region covered up until cuts were finalized, and
spans 130 $< \mgg <$ 140 MeV/c$^2$ and 
485 $< \meegg <$ 510 MeV/c$^2$.  The ellipse in the box is the signal
region, which spans $\sim$2$\sigma$ in the $\klpiee$ signal 
Monte Carlo $\meegg$ and $\mgg$ distributions.  In the $\meegg$ direction,
the ellipse is $\pm$ 5.02 MeV/c$^2$ wide, and in the $\mgg$ direction it is 
 $\pm$ 2.32 MeV/c$^2$ wide. 
The rectangular ``strip" to the left of the box is dominated by
backgrounds from $\ktwopid$ and $\kthreepid$ decays with accidental $\pi^0$s. 
Missing particles in these decays cause the reconstructed
mass $\meegg$ to be low.  Because these backgrounds accumulated in the strip,
this region is not considered in the background estimation described below.

In the background estimation, 
the data in Figure 1, outside the 
strip and box regions, is fit to the sum of  planar parts and 
the $\kleegg$ sample:\\
$f$ ($\meegg,\mgg$) = A$_0$ + A$_{\gamma\gamma}\mgg$~+ 
A$_{ee\gamma\gamma}\meegg$~+ \\
\hspace*{.3in}A$_g$ g($\meegg,\mgg$) \\
where
g($\meegg,\mgg$) is the $\kleegg$ distribution in the 
$\mgg$ vs. $\meegg$ plane.  The parameters $A_i$ were the parameters from the 
fit. The non-$\kleegg$ background was well-modeled with
first-order terms.  The estimated background in the signal ellipse is 
38.11 $\pm$ 1.67 events, with 0.27 $\pm$ 0.03 event contribution from
non $K_L \to ee\gamma \gamma$ backgrounds. 

In order to reduce this background, phase space cuts~\cite{greenlee}
are applied to the data.  The location of these cuts is optimized by
minimizing the expected 90\% C.L. branching ratio limit of $\klpiee$, 
using the Feldman and Cousins~\cite{feldman} methodology. The expected
branching ratio limit is computed by randomly generating a large ensemble
of virtual experiments in which the known background sources are the only
contributions to the observed number of events.  The effect of the  
cuts on signal efficiency is also accounted for. 

The phase space variables with the best discrimination against
$\kleegg$ background are $\mid y_{\gamma}\mid$ and $\thetamin$.  
The variable 
$y_{\gamma}$  is the cosine of the angle between the $\pi^0$ decay axis and 
the sum of the momenta of the two electrons, calculated in the center of 
mass of the photon pair. In the signal mode, $\mid y_{\gamma}\mid $ 
is nearly 
uniformly distributed because the pion has spin zero, but in $\kleegg$, the 
distribution is peaked at one. The variable $\thetamin$ is the minimum angle
between any photon and any electron in the kaon rest frame.  
It provides good separation because in $\kleegg$, a radiated photon 
typically has a small angle with respect to the electron from which it
originated, while in $\klpiee$, $\thetamin$ is nearly flat.
Distributions for $\mid y_{\gamma}\mid$ and $\thetamin$ 
in $\klpiee$ Monte Carlo 
and $\kleegg$ data and Monte Carlo appear in Figure 2.

\begin{figure}
\scalebox{0.9}{\includegraphics{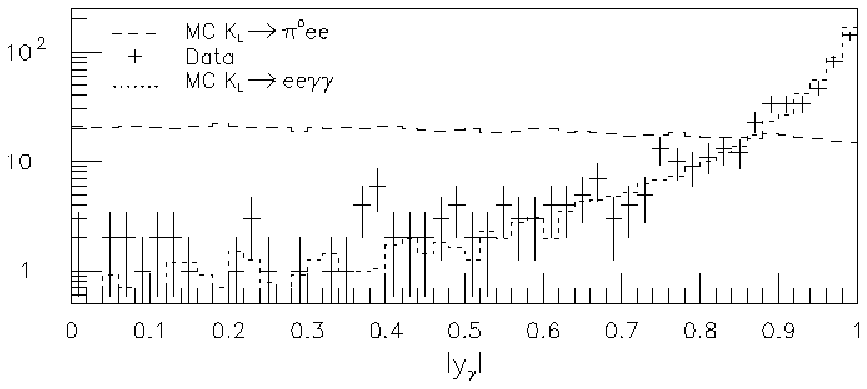}}
\scalebox{0.9}{\includegraphics{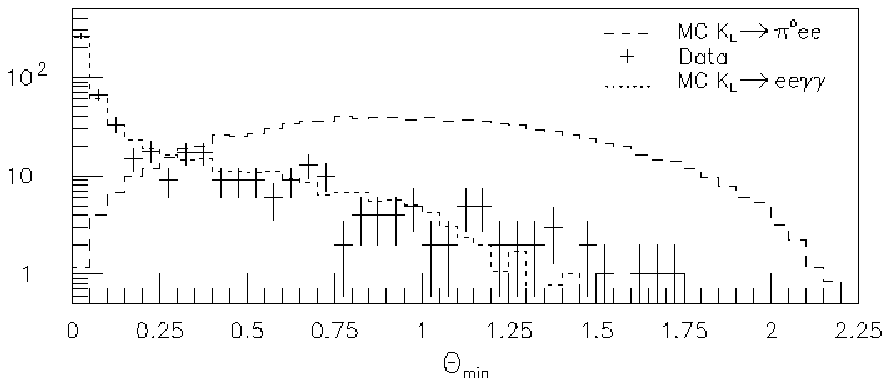}}
\caption{\label{fig:ygl_thetamin}
$\mid y_{\gamma}\mid $ (top) and $\thetamin$ (bottom)
distributions for $\klpiee$ MC and $\kleegg$ data and MC.
$\kleegg$ events come from inside the swath but outside the box.  
$\klpiee$ MC are from inside the box, and the normalization is arbitrary.}
\end{figure}

The optimized phase space cut values are $\thetamin > 0.362 \pm 0.017$
and $\mid y_{\gamma} \mid < 0.745 \pm 0.002$.  These cuts reduce the 
expected background from 38.11 $\pm$ 1.67 events to 0.99 
$\pm$ 0.35 
with a signal loss of 27\%.  The signal acceptance, assuming uniform 
three-body phase space, is (2.749 $\pm$ 0.013)\%,
giving a single event sensitivity of  1.04 $\times 10^{-10}$. 

\begin{figure}
\scalebox{1.0}{\includegraphics{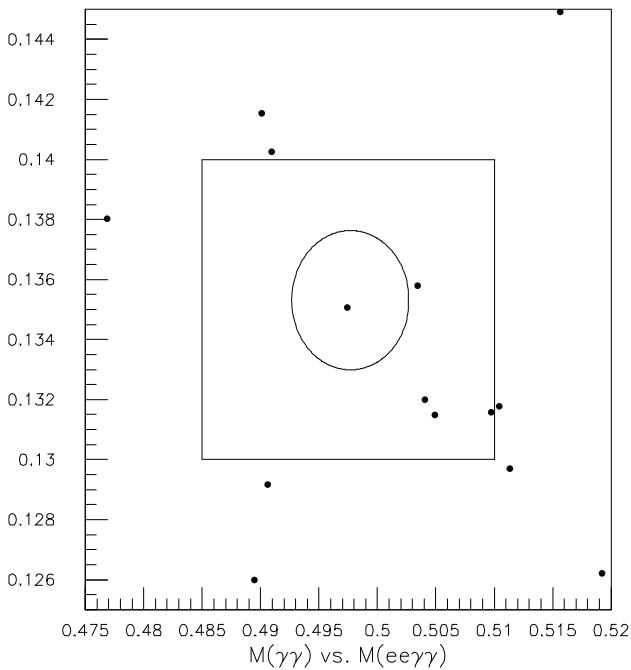}}
\caption{\label{fig:prl_zoom}
$\mgg$ vs. $\meegg$ in GeV/c$^2$ for the data after all cuts have been applied.
The box is open and one event appears within the signal ellipse, with a
background of 0.99 $\pm$ 0.35 events}
\end{figure}

When the box in Fig. 1 was opened (Fig. 3), one event was observed in
the signal ellipse.  Taking the background level into account, we determine
BR($\klpiee$) $<$ 3.50 $\times 10^{-10}$.  Combining this with the previous
result yields the final KTeV result:  BR($\klpiee$) $<$ 2.8 $\times 10^{-10}$
at 90\% C.L.

If instead of a uniform three-body phase space distribution for the signal 
mode, we assume a vector interaction model for the direct CPV part of the
decay and allow for form factors as in \cite{us}, we find for the combined 1997
and 1999 data samples an upper limit of BR($\klpiee$) 
$<$ 3.4 $\times 10^{-10}$.
If the decay $\klpiee$ is saturated by the direct
CPV component, we constrain the Wolfenstein CKM parameter
$|\eta_{CKM}| <$ 3.3.  Although other measurements yield a more
stringent constraint on $|\eta_{CKM}|$, it is important to make 
a variety of measurements
in both the kaon system and the B system to determine if the CKM parameters
are consistent. 

We gratefully acknowledge the support and effort of the Fermilab
staff and the technical staffs of the participating institutions for
their vital contributions.  This work was supported in part by the U.S.
Department of Energy, The National Science Foundation and The Ministry of
Education and Science of Japan.
In addition, A.R.B. and E.B.
acknowledge support from the NYI program of the NSF; A.R.B. and E.B. from
the Alfred P. Sloan Foundation; E.B. from the OJI program of the DOE;
K.H. from the Japan Society for the Promotion of
Science; and R.F.B. from the Funda\c{c}\~{a}o de Amparo \`{a}
Pesquisa do Estado de S\~{a}o Paolo.

\end{document}